\documentclass[
aps,
pra,
longbibliography,
superscriptaddress,
 amsmath,amssymb,
twocolumn,
]{revtex4-1}

\usepackage{graphicx}
\usepackage{dcolumn}
\usepackage{bm}
\usepackage{upgreek}
\usepackage{color}
\usepackage{hyperref}
\DeclareMathAlphabet{\mathpzc}{OT1}{pzc}{m}{it}
\usepackage{enumitem}   

\newcommand{\RR}{\right}
\newcommand{\LL}{\left}
\newcommand{\m}{\mathrm}
\newcommand{\dg}{\dagger}
\newcommand{\eref}[1]{Eq.~(\ref{#1})}
\newcommand{\fref}[1]{Fig.~\ref{#1}}
\newcommand{\puoli}{\frac{1}{2}}
\newcommand{\puolisqrt}{\frac{1}{\sqrt{2}}}

\usepackage[normalem]{ulem}

\setlist[enumerate]{itemsep=-1mm}

\begin{document}

\title{Quantum backaction evading measurements of a silicon nitride membrane resonator}

\author{Yulong Liu}
\affiliation{Beijing Academy of Quantum Information Sciences, Beijing 100193, China}
\affiliation{Department of Applied Physics, Aalto University, P.O. Box 15100, FI-00076 AALTO, Finland}

\author{Jingwei Zhou}
\affiliation{QTF Centre of Excellence, Department of Applied Physics, Aalto University, FI-00076 Aalto, Finland}

\author{Laure Mercier de L\'epinay}
\affiliation{QTF Centre of Excellence, Department of Applied Physics, Aalto University, FI-00076 Aalto, Finland}

\author{Mika A. Sillanp\"a\"a}
 \email{Mika.Sillanpaa@aalto.fi}
\affiliation{QTF Centre of Excellence, Department of Applied Physics, Aalto University, FI-00076 Aalto, Finland}

\date{\today}

\begin{abstract}
Quantum backaction disturbs the measurement of the position of a mechanical oscillator by introducing additional fluctuations. In a quantum backaction measurement technique, the backaction can be evaded, although at the cost of losing part of the information. In this work, we carry out such a quantum backaction measurement using a large 0.5 mm diameter silicon nitride membrane oscillator with 707 kHz frequency, via a microwave cavity readout. The measurement shows that quantum backaction noise can be evaded in the quadrature measurement of the motion of a large object. 
\end{abstract}

\maketitle


\section{Introduction}
Cavity optomechanics can be realized at microwave frequencies such that a conductive mechanical oscillator acts as a time-dependent capacitance, and modulates the resonance frequency of a GHz frequency microwave resonator. This creates an interaction analogous to radiation-pressure coupling realized with optical cavities and mechanical oscillators. Aluminum drum oscillators integrated with superconducting on-chip microwave resonators \cite{Teufel2011b} have been used in groundbreaking fundamental research on quantum-mechanical behavior of mechanical systems, see eg., Refs.~\cite{SchwabSqueeze,TeufelSqueeze,4BAE,Teufel2021entangle}.

Monitoring an external time-dependent force driving an oscillator has been a research topic of interest for several decades \cite{Caves1978QND,Caves1981qnoise}. Cavity optomechanics, especially microwave optomechanical systems, have been an excellent platform for investigating how fundamental limits influence the precision of these measurements. Not so long ago, experiments reached the state where quantum backaction of the measurement on the oscillator becomes observable \cite{RegalShot,Kippenberg2017Squ,Schliesser2018sql,Corbitt2019SQL,LIGO2020squeeze}. In some measurements conducted at deep cryogenic temperatures, quantum backaction heating of the oscillator has been the dominant noise source \cite{Teufel2016ShotN}.

As per quantum mechanics, measuring a system inevitably disturbs it. The system collapses to an eigenstate of the measured observable, leaving the conjugate observable undefined. 
Some measurement techniques, however, rely on coupling a probe to the system in such a way that the measured observable is an invariant of the Hamiltonian evolution, which allows to preserve the state of at least this observable. These are often called quantum nondemolition (QND) measurements. Such measurements have been demonstrated for observables of simple, relatively pure systems such as single photons \cite{Haroche1999QND}, or electrical degrees of freedom in superconducting qubits \cite{DelftQND07,Siddiqi2011}. However, they remain technically difficult to achieve in macroscopic systems such as massive mechanical oscillators involved in force measurement, which have more chances of coupling to their environment in uncontrolled ways, restoring possibilities for state demolition.

When a mechanical oscillator's position is continuously monitored via a cavity, the measurement is not typically QND even ideally. The ``collapse" in the continuous position measurement implies quantum backaction disturbance to momentum, however, the disturbance will leak also into the position in the process. Thus, this measurement is not QND. One could search for ways of measuring the oscillator's energy rather than position, which would conserve the oscillator's state.
However, no real alternative applicable to large-scale interferometers, which are a significant beneficiary of measurements beyond conventional quantum limits, have been found.

An approach for conducting QND quadrature measurements via a cavity, however, is available in optomechanics \cite{Marquardt2008Sq} and has been investigated in several experiments \cite{Schwab2010QND,Schwab2012instab,Schwab2014QND,TeufelSqueeze,Schwab2016squ3db,2BAE,Entanglement,Kippenberg2019BAE,4BAE}. In this context, the term quantum backaction evading (BAE) measurement has been used instead of QND. The idea is to modulate the cavity field amplitude synchronously to the mechanical vibrations. As a result, the cavity field couples only to one quadrature of the oscillation, allowing to measure this quadrature in a BAE manner.

In this work, we extend the scope of BAE measurements to a new class of systems with a high degree of coherence and therefore immediately adapted to force or metric sensing. As the mechanical oscillator, we use a silicon nitride membrane embedded in a microwave cavity \cite{Steele3D,Steele2015HighQ,Nakamura3D,Schliesser2021SiN}. High-stress silicon nitride (SiN) has emerged as the material to realize the highest mechanical quality factors for usage in quantum optomechanics \cite{Regal2009SiN,Weig2011,Harris2015asym,Groblacher2016,Regal2017,Schliesser2018FB,Schliesser2018sql,Kippenberg2018dissip}. This is due to the fact that a high pre-stress shunts the material losses, and the mode profile can be designed such that anchor losses are mitigated. In our work, a high $Q$ value assists in reaching conditions where quantum backaction becomes relevant. A relatively low frequency and narrow spectral linewidth, however, create experimental challenges, which we show can be overcome.

\begin{figure}[h]
  \begin{center}
   {\includegraphics[width=0.75\columnwidth]{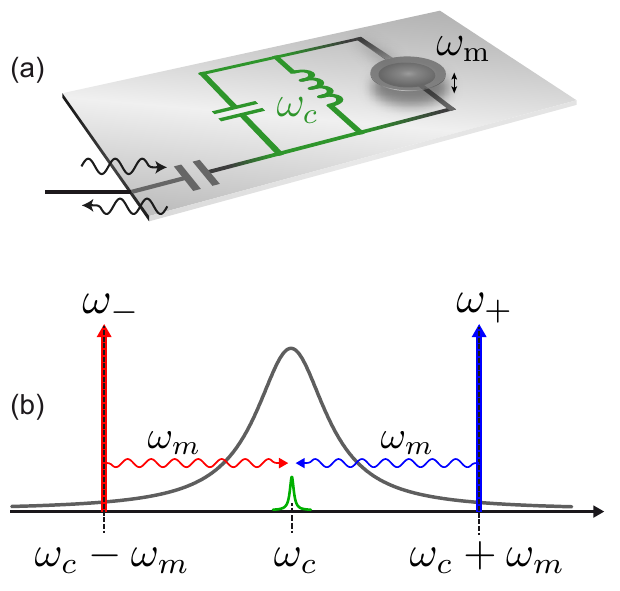} }
    \caption{\textit{Principle of quantum backaction evading (BAE) measurements in cavity optomechanics.} (a) In a microwave optomechanical cavity, a metallized membrane forms a vibrating plate capacitor together with a counter electrode. (b) The frequency scheme shows the cavity mode being pumped with equal tones at red and blue sideband frequencies. The green mechanical peak in center illustrates the preferential scattering of the pump tones.}
    \label{fig:scheme1}
 \end{center}
\end{figure}

\section{Position measurement through an optomechanical cavity}

Let us recall how the position measurement can be carried out in cavity optomechanics. A mechanical oscillator is assumed to modulate the resonance frequency of a cavity. The latter is typically an optical cavity, but the resonance can equally well be in the microwave-frequency range \cite{Lehnert2008Nph}. In microwave optomechanics, the system is readily understood as depicted in \fref{fig:scheme1} (a). The membrane acts as a vibrating plate capacitor in a microwave circuit (capacitance $C$ and inductance $L$), thus modulating $C$ and the resonance frequency of the cavity. 

We describe the cavity with the annihilation operator $a$, and the oscillator correspondingly with $b$. For the oscillator, the dimensionless position and momenta are
\begin{equation}
\begin{split}
\label{eq:mecpos}
 x(t) & = \frac{1}{\sqrt{2}}(b^\dg + b) \,, \\
 p(t) & = \frac{i}{\sqrt{2}}(b^\dg - b)\,.
\end{split}
 \end{equation}
Similarly, the scaled currents and voltages in the cavity are
\begin{equation}
\begin{split}
\label{eq:cavpos}
 x_c(t) & = \frac{1}{\sqrt{2}}(a^\dg + a) \,, \\
 p_c(t) & = \frac{i}{\sqrt{2}}(a^\dg - a)\,.
\end{split}
 \end{equation}

The Hamiltonian of the system is
\begin{equation}
\label{eq:H}
H = \frac{\omega_c}{2} \left(x_c^2 + p_c^2 \right) + \frac{\omega_m}{2} \left(x^2 + p^2\right) + \sqrt{2} g_0 x_c^2 x \,.
\end{equation}
%
%
%
Here, $\omega_c$ is the resonance frequency of the cavity, and $\omega_m$ is that of the mechanical oscillator. The cavity and the oscillator are further characterized by the damping rates $\kappa$ and $\gamma$. The cavity losses are additionally divided into external (e) and internal (i) losses according to $\kappa = \kappa_e +\kappa_i$. The optomechanical coupling coefficient in frequency units is defined as 
\begin{equation}
\begin{split}
\label{eq:g0}
g_0 = \frac{d \omega_c}{dx} = \frac{\omega_c}{2 C} \frac{d C}{dx} \,,
\end{split}
 \end{equation}
where the latter form holds for the microwave-cavity version.

The above dynamical variables are defined in the laboratory frame, that is, in the Schr\"odinger picture. They are denoted by lowercase letters. Below, the variables are predominantly defined in the interaction picture, which are denoted in uppercase and in \eref{eq:sch-heis} with a $I$ subscript. The position, as an example, can be written with the slowly varying quadrature amplitudes $X_I$ and $P_I$, defined in a rotating frame set by a frequency $\omega_x$, according to
\begin{equation}
\begin{split}
\label{eq:sch-heis}
x(t) &= X_I(t) \cos \omega_x t - P_I(t) \sin \omega_x t  \\
& = \puolisqrt \LL[ b_{I} \exp (i\omega_x t) + b_{I}^\dg  \exp (-i\omega_x t) \RR] \,.
\end{split}
 \end{equation}

 %
Below, $\omega_x$ can be the frequency of either of the individual harmonic modes, or that of the pump tone to the cavity. For simplicity, we omit the $I$-subscripts in the following.

There are several specific ways of realizing the position measurements, which differ in terms of pumping or the parameter regime, which are discussed next. All of them, of course, exhibit QBA on the oscillator. In the simplest case, the optomechanical system is pumped by a single laser or microwave tone. This situation can be further separated into unresolved or resolved-sideband cases. The discussion is presented because when analyzing the experimental results, our main interest is to find out if QBA is evaded. Identifying an exact comparison point to answer this question is not straightforward and remains somewhat ambiguous, because there is no model-independent way to do either position measurement, or BAE quadrature measurement. Nonetheless, our aim is to make a worst-case comparison of the experimentally measured backaction to the different theoretical expectations.

\subsection{Direct position measurement}

In the basic optomechanical measurement, a single coherent pump tone is applied to the cavity at the frequency $\omega_d$, inducing a pump photon number $n_c$. The pump detuning is defined as $\Delta = \omega_d - \omega_c$. 
We also define the susceptibilities of the mechanics and of the cavity as $\chi_m^{-1} = \frac{\gamma}{2} - i \omega$ and $\chi_c^{-1} = \frac{\kappa}{2} - i \omega$. 

We transfer the cavity operators to the rotating frame at the frequency $\omega_d$ of the pump (e.g.,~$x_c \rightarrow X_c$, while the mechanics stays in the lab frame). Under a strong pumping of, say the $X_c$ quadrature of the cavity (making a choice of phase), the pumping induces a cavity field $X_c \rightarrow \bar{X_c} + X_c$, where $\bar{X_c}$ is the large coherent field and $X_c$ describes small fluctuations. With this change to \eref{eq:H}, the system is linearized as
\begin{equation}
H = -\frac{\Delta}{2} \left(X_c^2 + P_c^2 \right) + \frac{\omega_m}{2} \left(x^2 + p^2\right) + 2 G X_c x \,,
\label{eq:Hrotcav}
\end{equation}
thus also obtaining the enhanced (effective) optomechanical coupling
\begin{equation}
G = g_0 \sqrt{n_c} \,.
\label{eq:G}
\end{equation}
The strength of interaction is also conveniently expressed as the cooperativity
\begin{equation}
C = \frac{4G^2}{\kappa \gamma} \,.
\label{eq:C}
\end{equation}

The input-output theory equations of motion in the frequency domain become
\begin{subequations}
\label{eq:eqmotXP}
\begin{alignat}{2}
  &\chi_c^{-1} X_c && = - \Delta P_c + \sqrt{\kappa_e} X_\mathrm{c,in} + \sqrt{\kappa_i} X_\mathrm{I,in} \,, \\
  &\chi_c^{-1} P_c &&= \Delta X_c- 2 G x + \sqrt{\kappa_e} P_\mathrm{c,in} + \sqrt{\kappa_i} P_\mathrm{I,in} \,, \\
  &\chi_m^{-1} x &&= \omega_{\mathrm{m}} p + \sqrt{\gamma} x_\mathrm{in} \,,\\
   &\chi_m^{-1} p &&= - \omega_{\mathrm{m}} x- 2 G X_c + \sqrt{\gamma} p_\mathrm{in} \,.
\end{alignat}
\end{subequations}
%
The input noise operators associated to the losses have the white spectral densities $\langle z(\omega) z(-\omega) \rangle = \puoli + n_z^T$ for all $z = X_\mathrm{c,in}, P_\mathrm{c,in}, X_\mathrm{I,in}, P_\mathrm{I,in}, x_\mathrm{in}, p_\mathrm{in}$. $n_z^T$ is the thermal occupation of a given bath, in particular $n_m^T = \left[ \exp (\hbar \omega_m /k_B T) - 1\right]^{-1}$ is the occupation of the mechanical oscillator at the temperature $T$. The occupation number relates to the variances of the dimensionless position and momentum as $n_m^T + \puoli = \langle b^\dg b \rangle + \puoli = \puoli \langle x^2 \rangle + \puoli \langle p^2 \rangle$, and similarly for the quadrature amplitudes. We assume the cavity external bath is at zero temperature, while there can be internal heating of the cavity described by $n_I^T$. The cavity heating is most conveniently given by the cavity thermal occupation as $n_c^T = n_I^T \frac{\kappa_i}{\kappa}$.

\subsubsection{Unresolved sidebands}

The unresolved sidebands situation $\kappa \gg \omega_m$ is a basic model for a position measurement. A fast cavity reacts to the oscillator without a time delay, and $\chi_c = 2/\kappa$. In Eqs.~(\ref{eq:eqmotXP}), we also assume $\Delta = 0$. The position $x$ is therefore measured by the cavity $P_c$ quadrature. The oscillator momentum suffers a backaction from $X_c$, which further leaks into $x$ in the dynamics. As a consequence, the backaction affects position and momentum in the same manner, given by the variances in units of quanta:
\begin{equation}
\label{eq:xpnoBAE}
\begin{split}
& \langle x^2 \rangle =\langle p^2 \rangle = \puoli + n_m^T + C(1+2n_c^T) \,.
\end{split}
\end{equation}
The terms on the rhs correspond, respectively, to vacuum, thermal, and backaction noise, with the quantum contribution to the backaction given by
\begin{equation}
\label{eq:qbanoBAE}
\langle x^2 \rangle_{qba} = \langle p^2 \rangle_{qba} = n_{qba}= C \,.
\end{equation}
On top of the quantum backaction, the term $2C n_c^T$ in \eref{eq:xpnoBAE} is an additional backaction due to classical cavity noise.


\subsubsection{Resolved sidebands}
\label{sec:red}

We now treat the opposite case, $\kappa \ll \omega_m$. Here, the cavity reacts only slowly to the mechanics, and the dominant consequence becomes the dynamical backaction on the mechanical oscillator. In particular, this implies the possibility to carry out sideband cooling of the mechanics down to the ground state when the pump detuning is set at the red sideband, viz., $\Delta = -\omega_m$.

\begin{equation}
\label{eq:xvarGoodCav}
\begin{split}
& \langle x^2 \rangle = \langle p^2 \rangle =  \frac{ 1 }{ 1+C} \LL[\puoli + n_m^T + C \LL(\puoli + n_c^T \RR) \RR] \,.
\end{split}
\end{equation}
The term $\LL(\puoli + n_m^T \RR)(1+C)^{-1}$ is the result of the sideband cooling process, where the effective mechanical damping is the sum of the intrinsic damping, and of the "optical" damping $\gamma_{\m{opt}}$ that is linear in pumping power:
\begin{subequations}
\begin{alignat}{2}
\gamma_{\m{eff}} & = \gamma + \gamma_{\m{opt}} \,, \label{eq:gammaeff} \\
\gamma_{\m{opt}} & = \frac{4 G^2}{ \kappa} \,. 
\label{eq:gammaopt}
\end{alignat}
\end{subequations}
When the sideband cooling is started from a high temperature, the final occupation number of the mechanical oscillator becomes
\begin{equation}
\label{eq:nmSBcool}
\begin{split}
n_m \approx \frac{\gamma }{\gamma_{\m{eff}}}n_m^T + n_c^T \,.
\end{split}
\end{equation}

Backaction is generally associated with heating, however, a quantum backaction is also associated with the measurement under sideband cooling conditions, although an overall effect is a strong cooling. One can consider the measurement as a measurement of an effective number of phonons $n_{\rm eff} = \frac{1}{C+1} \left(\puoli + n_m^T\right)$. Cavity quantum noise generates an increased effective number of phonons:
\begin{equation}
\langle x^2 \rangle_{qba, \rm eff} =  \frac{C}{2(1+C)} \,,
\end{equation}
and the effect of the quantum backaction on the measured energy of the initial bath (independently from dynamical backaction) is:
\begin{equation}
\label{eq:qbaSBcool}
\langle x^2 \rangle_{qba}= n_{qba} = \frac{C}{2} \,.
\end{equation}
%
The reason for only a half the backaction as compared to \eref{eq:qbanoBAE} is that in \eref{eq:qbanoBAE}, the backaction comes from each of the two frequency bands of the cavity field at which the spectrum is projected.

\subsection{Single-quadrature BAE measurement}

Position cannot be measured in a continuous and QND manner, but the quadrature amplitudes, as defined in \eref{eq:sch-heis}, can be. The idea is to modulate the interaction between the oscillator and the detector (cavity in this case) in synchrony with the motion \cite{BraginskyQND,Onafrio1996force,Marquardt2008Sq}. In optomechanics this is accomplished as depicted in \fref{fig:scheme1} (b). Two equal-power coherent tones are applied exactly at the mechanical sidebands, at the frequencies $\omega_\pm = \omega_c \pm \omega_m$. The complex amplitudes of the fields at these frequencies are denoted by $|\alpha| \exp(i\theta_\pm)$. We move the cavity operators to a frame rotating at the mean frequency of the two pumps (that is, the cavity frequency), and the mechanics to a frame rotating at $\omega_m$. In this frame, we define the phase difference of the tones as $\theta = (\theta_+ - \theta_-)/2$. After linearization, and supposing a good-cavity situation $\kappa \ll \omega_m$, which allows us to neglect terms oscillating at $\pm 2\omega_m$, \eref{eq:H} becomes
\begin{equation}
\label{eq:hBAE}
\begin{split}
&H=  |G| \LL[ \exp(i \theta) a^\dg  + \exp(-i \theta)a  \RR] \LL(b^\dg + b \RR) = 2 G X_c^\theta X \,,
 \end{split}
 \end{equation}
where the rotated cavity quadrature is $X_c^\theta = X_c \cos \theta + Y_c \sin \theta$. Notably, all the dynamical variables are now  quadrature amplitudes. The effective coupling is $G = g_0 |\alpha|$. The mechanical quadrature $X$ is now a constant of motion since $[X,H] = 0$, while the unmeasured $P$ quadrature absorbs all the backaction. 

We also consider a finite noise energy in the cavity, which is characterized by the cavity thermal occupation $n_c^T$. We denote the spectral densities with the symbol $S[\omega]$, for example, the spectrum of the quadrature $X$ is $S_X[\omega] = \langle X [\omega] X[-\omega] \rangle$. The mechanical quadrature spectra become
\begin{equation}
\label{eq:BAESQSP}
\begin{split}
 S_X[\omega] &
= \frac{\gamma}{\omega^2 + (\gamma/2)^2} \left( \puoli + n_m^T\right)  \,,\\
 S_P[\omega]
&=  S_X[\omega] + \frac{16 G^2}{\left[\omega^2 + (\gamma/2)^2\right] \kappa}  \left( \puoli + n_c^T\right) \,.
\end{split}
\end{equation}
In addition to quantum backaction, any cavity noise $n_c^T$ heats up the $P$ quadrature.

The total number of quanta in the mechanical quadratures is 
\begin{equation}
\label{eq:XPvarBAE}
\begin{split}
& \langle X^2 \rangle  =\puoli + n_m^T \,,\\
& \langle P^2 \rangle =  \puoli + n_m^T + 2C+ 4C  n_c^T \,.
\end{split}
\end{equation}
The backaction is hence entirely projected onto the $P$ quadrature:
\begin{equation}
\label{eq:qbaBAE}
\begin{split}
\langle P^2 \rangle_{qba} & = 2C \,, \\
n_{qba} & = C \,.
\end{split}
\end{equation}

The quantum backaction on the quadratures at a given cooperativity is thus different in all the three example models, Eqs.~(\ref{eq:qbanoBAE},\ref{eq:qbaSBcool},\ref{eq:qbaBAE}). The total backaction on the oscillator is the same in Eqs.~(\ref{eq:qbanoBAE},\ref{eq:qbaBAE}), although this can be seen as incidental. The backaction in the two-tone BAE model is most closely related to the resolved-sideband case discussed in Section \ref{sec:red}. In the BAE case, at a given cooperativity, there is double the power, and thus a double total backaction.


In experiment, the signal is in the incoherently emitted field 
from the cavity. In microwave measurements, one measures the spectrum of the voltage fluctuations (output spectrum $S_\m{out}[\omega]$), which is obtained from the standard input-output theory:
\begin{equation}
\label{eq:SoutBAE}
\begin{split}
& S_\m{out}[\omega] = \frac{4C  \gamma \kappa_e}{\kappa} S_X[\omega] 
+ \frac{4 \kappa_e}{\kappa} n_c^T +  S_\m{in}[\omega] \,. 
\end{split}
\end{equation}
Here, $S_\m{in}[\omega] = \langle P_{c,in}[\omega] P_{c,in}[-\omega] \rangle$ is the incoming cavity noise, which in the experiment is assumed to be the vacuum noise, satisfying $S_\m{in}[\omega] = \puoli$. The actual measured spectra contains also a white noise contribution by the noise added by the amplifier. The relevant information, the Lorentzian given by the first term on the rhs in \eref{eq:SoutBAE}, however, sits on top of any noise floor and is easily recognized. In the following, we also use the integrated mechanical spectrum, provided by this peak:
\begin{equation}
\label{eq:noutBAE}
\begin{split}
n_\m{out} = \frac{4C  \gamma \kappa_e}{\kappa} \langle X^2 \rangle \,. 
\end{split}
\end{equation}

We also note variants of the single-quadrature detection idea, where collective motion of two oscillators can be measured in BAE fashion \cite{Caves2012BAE,WoolleyBAE,2BAE,Polzik2017,4BAE}.

\section{Experiment}

\subsection{Device}

The oscillating membrane is a standard 0.5 mm square window from Norcada, made of 50 nm thick SiN with a high pre-stress of 900 MPa. It is metallized in the center by 20 nm Al in the form of a 200 $\mu$m square.

\begin{figure}[h]
  \begin{center}
   {\includegraphics[width=0.99\columnwidth]{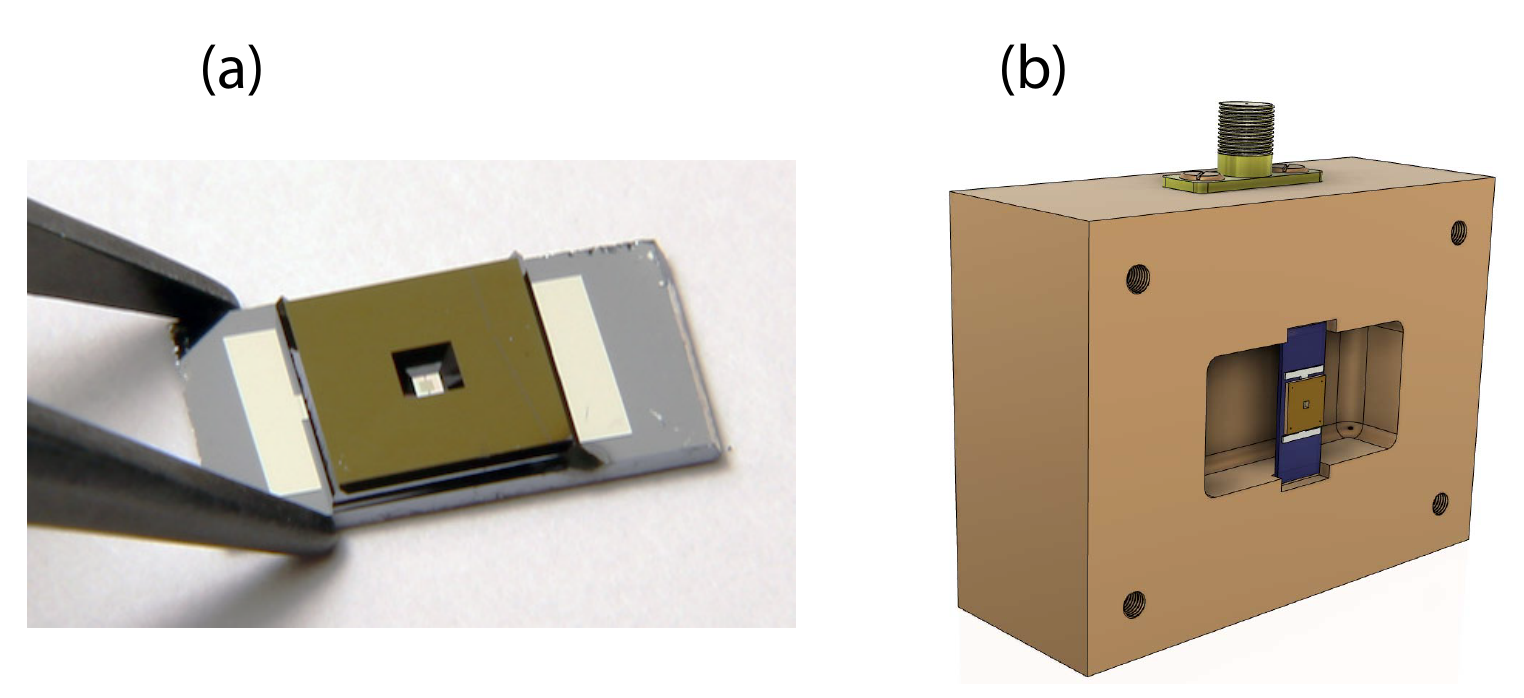} }
    \caption{\textit{Sample design.} (a) Photograph of a device similar to the measured one. A membrane chip is flip-chipped on top of a larger antenna chip. (b) The chip assembly inside a 3-dimensional microwave cavity made of annealed high-conductivity copper. The rendering shows a lower half of the cavity.}
    \label{fig:sample}
 \end{center}
\end{figure}

For coupling to microwave mode, we use flip-chip technique similar to \cite{Steele3D,Steele2015HighQ,Gravity2021}. Antenna pattern, with 120 nm of Al metallization, is realized on a separate high-purity Si chip. The membrane chip is glued on the antenna chip from one corner using a drop of fast epoxy [see \fref{fig:sample} (a)]. Without any support posts between the chips, or removal of Si from the antenna chip outside the metallization, the natural gap between the chips becomes on the order 500 nm. The gap is determined by irregularities, or possibly bending of the membrane chip resulting from the fact that the stressed SiN is removed from the backside. The dimensions yield an estimated $g_0/2\pi \simeq 10$ Hz.

As depicted in \fref{fig:sample} (b), the chip assembly is embedded in a 3D microwave cavity \cite{Schoelkopf3D} made of oxygen-free copper that is annealed at $850^{\circ}$C for 8 hours, with a small air flow at $2\times 10^{-3}$ mBar. We found that (non-superconducting) Cu cavities provided better thermalization of the mechanics in comparison to superconducting Al cavities, where the mode usually does not thermalize below $\sim 100$ mK. The reason is likely bad thermal conductivity of the superconductor.

Because the membranes are highly sensitive to vibration noise of the pulse tube in contemporary cryogen-free dilution refrigerators, the measurements are conducted in a wet liquid-He based dilution refrigerator, which has a base temperature of about 25 mK.

The mechanical mode (the lowest flexural mode) has the frequency $\omega_m/2\pi \simeq 707.4$ kHz. Its damping rate measured with ringdown is $\gamma/2\pi \simeq 8.8$ mHz, corresponding to quality factor $Q_m \simeq 80$ millions. The cavity resonance is found at $\omega_c/2\pi \simeq 4.517$ GHz. At temperatures below around 300 mK, the cavity has the internal and external loss rates $\kappa_i/2\pi \simeq 156$ kHz, and $\kappa_e/2\pi \simeq 145$ kHz, which sum up to $\kappa/2\pi \simeq 301$ kHz. 

At room temperature, the Cu cavities were found to have a $Q$ of 2000-4000. We expect that well-annealed Cu exhibits a bulk conductivity enhancement of 2-3 orders of magnitude when cooled down below 4 K. The observed increase of $Q$ is much less than anticipated from the enhanced conductivity. We therefore believe that conductive surface oxides limit the $Q$. Nonetheless, even though the cavity is not superconducting, we can approach the good-cavity situation $\omega_m \gg \kappa$.

\subsection{Single-tone measurements}

\begin{figure}[h]
  \begin{center}
   {\includegraphics[width=0.85\columnwidth]{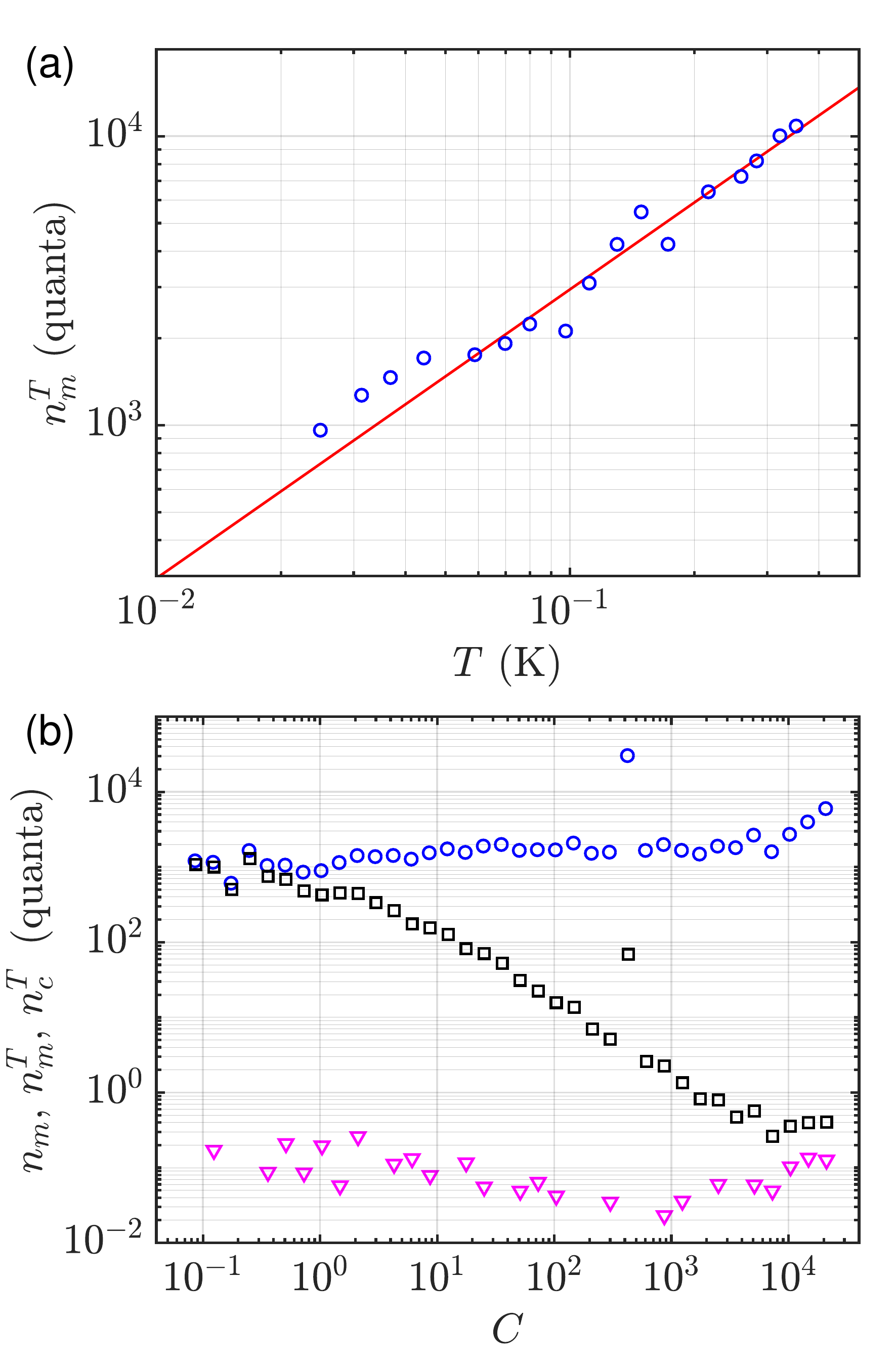} }
    \caption{\textit{Thermalization and sideband cooling.} (a) Mechanical phonon number as a function of the cryostat temperature. The straight line is a linear fit. (b) Ground-state cooling starts from the equilibrium mode temperature of $37$ mK. The graph shows the relevant occupation numbers as a function of cooperativity. Squares: $n_m$; circles: $n_m^T$; triangles: $n_c^T$.  
    }
    \label{fig:Tsweep}
 \end{center}
\end{figure}

Under red-sideband measurement, the mechanical mode temperature is proportional to the area of the recorded mechanical spectral peak. In cryogenic optomechanics experiments, the mechanical mode may be excited to energies higher than expected based on the refrigerator temperature. This is associated to low thermal conductivity, (technical) heating by the pump irradiation, or external vibration noise. In the present case, the mechanical mode is found to thermalize rather well, down to $30...40$ mK when the cryostat temperature is varied, as seen in \fref{fig:Tsweep} (a). An important calibration point in the ensuing analysis is given by the equilibrium mode occupation $n_m^{T,0}$ at a given refrigerator temperature. In the forthcoming experiments, we operate at the base temperature of 25 mK. We determine $n_m^{T,0}$ as the average of the two lowest-temperature data points in \fref{fig:Tsweep}, obtaining $n_m^{T,0} \simeq 1100$. The rest of the calibration procedure is detailed in Appendix \ref{appendix}.

Standard sideband cooling is performed by irradiating the cavity with a strong red-sideband tone. This measurement is used to benchmark the performance of the device. It also allows to calibrate mechanical parameters such as the effective mechanical damping rate. 

As depicted in \eref{eq:nmSBcool}, sideband cooling allows in the limit of large effective damping $\gamma_{\rm eff}$, cooling the mechanical mode close to the ground state where $n_m = 0$. This has been demonstrated in many different types of mechanical oscillators, including SiN membranes \cite{Nakamura3D,Schliesser2021SiN}.

We show sideband cooling close to the ground-state in \fref{fig:Tsweep} (b), reaching $n_m \simeq 0.3$ at $C \simeq 10^4$ where $\gamma_{\m{opt}}/2\pi \simeq 90$ Hz, with an outlier data point probably due to a building vibration. At higher powers, we observe significant technical heating of the mechanical bath, displayed also in \fref{fig:Tsweep} (b). This heating is seen in many microwave optomechanical experiments \cite{Lehnert2008Nph, SchwabSqueeze,Squeeze,Entanglement,4BAE}. The exact mechanism is unknown. Technical heating of the cavity mode, however, is nearly absent as seen in \fref{fig:Tsweep} (b). This is in stark contrast to results obtained with samples we measured in superconducting Al cavities, where the cavity mode can heat up to tens of quanta. This hints that the poor thermal conductivity of superconductors has a role in cavity heating.

\subsection{Two-tone BAE measurements}

\begin{figure}[h]
  \begin{center}
   {\includegraphics[width=0.85\columnwidth]{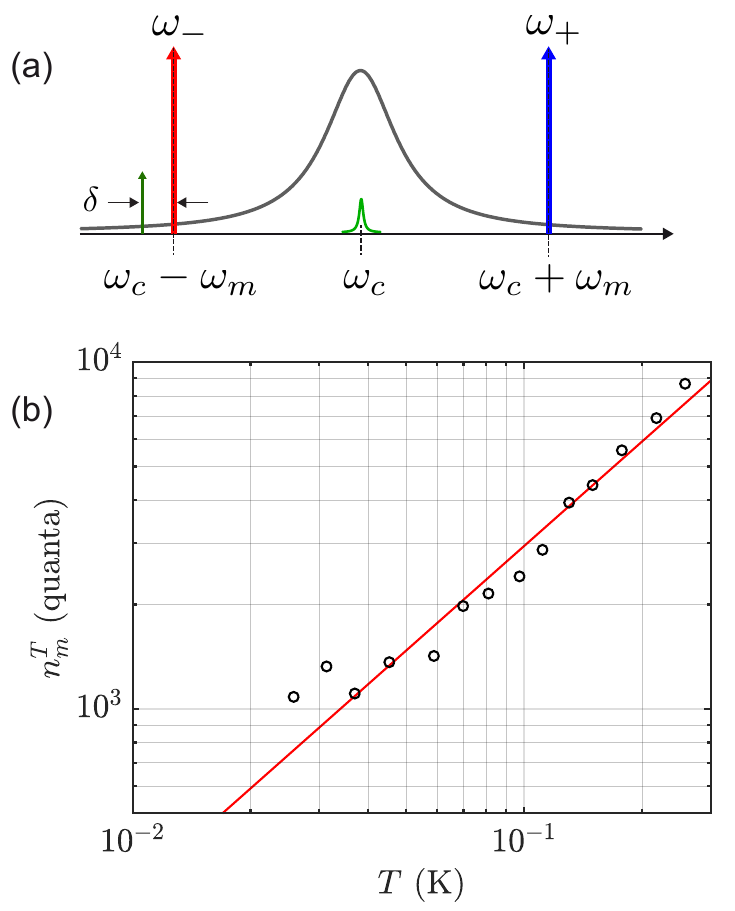} }
    \caption{\textit{Temperature sweep in BAE-measurement.} (a) Pump scheme. An additional strong sideband cooling tone is applied to broaden the mechanical linewidth up to $\gamma_\m{eff}/2\pi \simeq 3.1$ Hz. (b) Bath phonon number as a function of temperature inferred from the mechanical spectrum. The BAE cooperativity was $C \simeq 2$, and cooling tone detuning $\delta/2\pi = 400$ Hz.}
    \label{fig:TsweepBAE}
 \end{center}
\end{figure}

To properly reach quantum backaction evasion using the scheme in \fref{fig:scheme1} (b), the frequency separation of the pump tones need to equal twice the mechanical frequency within at least a few tens of percent of the mechanical linewidth. The mHz intrinsic mechanical linewidth of the membranes makes this nearly unreachable, especially as the mechanical frequency was observed to jump intermittently at the scale of $\gamma$. Therefore, we use an additional cooling tone [see \fref{fig:TsweepBAE} (a)]. The cooling tone cooperativity was chosen such that we obtain $\gamma_\m{eff}$ on the order one Hz, which relieves the above problem. The frequency of the cooling tone was incommensurately spaced from the BAE tones (detuned from the red sideband by $\delta = 400$ Hz in the experiment), and thus the cooling process can be treated independently of the BAE process. In terms of the mechanical spectrum, \eref{eq:BAESQSP}, the bath temperature becomes replaced by the mode temperature, viz.~$n_m^T \rightarrow n_m$ [\eref{eq:nmSBcool}].

Same as with sideband cooling, we need a calibration point for $n_m^T$ (or $n_m$). Although under purely red-detuned tone irradiation, thermalization and technical heating were very favorable, this may not hold true under more complicated driving. The calibration is obtained by performing a temperature sweep as shown in \fref{fig:TsweepBAE}(b). The integrated mechanical peak in the output spectrum gives the energy of the mechanical mode as per \eref{eq:noutBAE}. There are non-characterized amounts of both gain and attenuation in the output cabling, and thus we cannot obtain absolute calibration of the mode temperature using \eref{eq:noutBAE}. We assume a good thermalization at several hundred mK temperatures, and use a linear fit of the peak area in this region as the basis of the calibrating $n_m^T$ (and $n_m$) at the base temperature where the final measurements are performed. The bath temperature is finally inferred from \eref{eq:nmTBAEcal}.

In the measurement, we compensated for the temperature dependence of the mechanical frequency by correcting the pump tone frequencies based on a calibration of $\omega_m$ at each temperature. As seen in \fref{fig:TsweepBAE} (b), the thermalization in the BAE scheme is comparable to that with a single-tone pumping, only with a modest thermal decoupling close to the base temperature.

\begin{figure*}[t]
  \begin{center}
   {\includegraphics[width=0.95\textwidth]{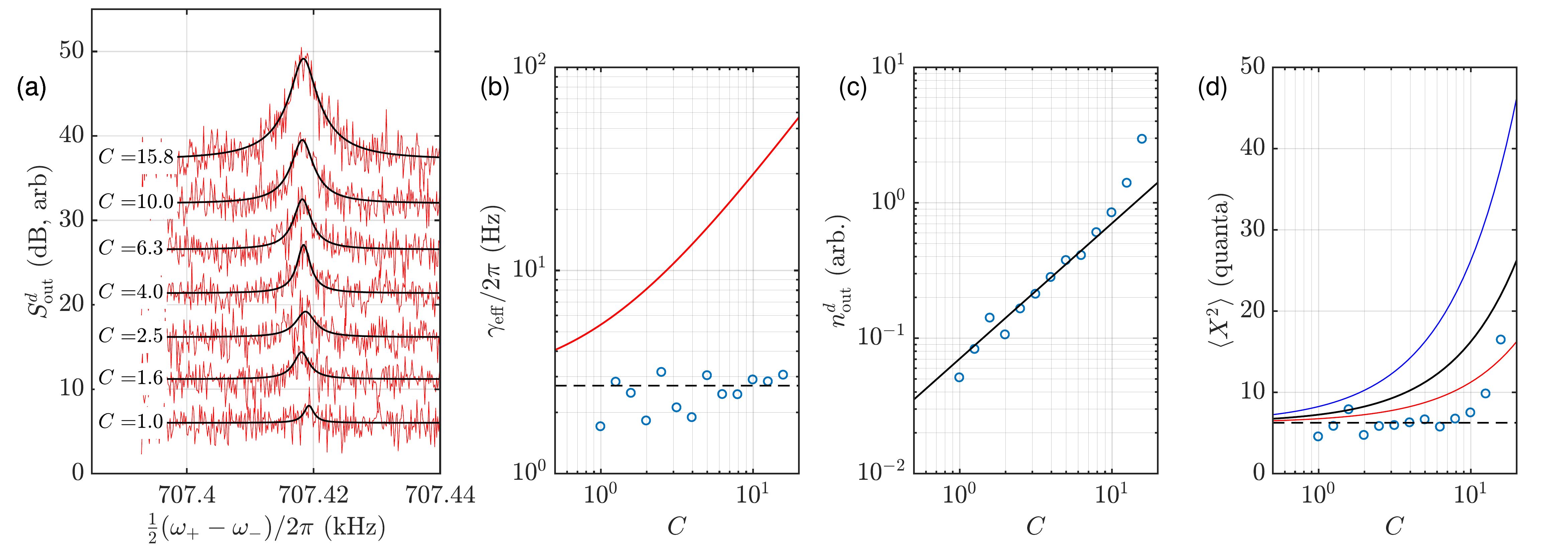} }
    \caption{\textit{BAE measurement.} (a) Output spectra at increasing cooperativities as indicated in the panel. The curves are shifted vertically for clarity. (b) Linewidths of the spectra. The dashed line marks the intrinsic linewidth (after broadening by the auxiliary cooling tone), while the solid line illustrates an expected broadening by red-only pumping. (c) Photon flux obtained from the integrated spectra. (d) Phonon number of the oscillator. The dashed line illustrates ideal BAE conditions. The solid lines include the contribution of quantum backaction in the three model cases considered; Solid black line: QBA on $x$, single-tone, bad cavity [\eref{eq:qbanoBAE}]; Solid red line: QBA on $x$, single-tone, good-cavity [\eref{eq:qbaSBcool}]; Solid blue line: QBA on $P$ quadrature in BAE [\eref{eq:qbaBAE}].}
    \label{fig:BAEmaindata}
 \end{center}
\end{figure*}

We now describe the BAE measurement at the base temperature, aiming at showing evasion of QBA that would appear as an increased occupation of the mechanics. Like in the temperature sweep, we used a strong cooling tone broadening the linewidth, here up to $\gamma_\m{eff}/2\pi \simeq 2.9$ Hz. In \fref{fig:BAEmaindata}, we first display in panel (a) the detected output spectra at different cooperativities, together with Lorentzian fits. An important benchmarking of BAE is that it does not influence the spectrum, including linewidth, of the measured $X$ quadrature. This is verified in \fref{fig:BAEmaindata} (b), where the spectral broadening resulting from the additional sideband cooling is also shown as a reference. In spite of some scatter at low powers, the linewidth is essentially unaffected by a strong BAE measurement. We then proceed to presenting the output flux in \fref{fig:BAEmaindata} (c) [see Eqs.~(\ref{eq:noutBAE},\ref{eq:BAEcalResults})]. The flux first grows linearly with power, but then overshoots the linear trend, which is associated to the technical heating of the mechanics. 

Finally, in \fref{fig:BAEmaindata} (d) we display the main result. Using the calibration procedure summarized in \eref{eq:BAEcalResults}, we obtain the fluctuations of the measured $X$ quadrature. The overshoot seen above in panel (c) now appears as heating of  $X$ at cooperativities $C \gtrsim 10$. As discussed above, there is no clear model-independent way to define what an expected QBA in an optomechanical measurement should be. When treating the two-tone BAE measurement, the difficulty is that the cooperativity is not in direct correspondence to the cooperativity in single-tone measurements. A possible comparison point comes from the QBA on the $P$ quadrature, equaling $2C$, in the BAE measurement. If this would be equally distributed between the two quadratures, each would receive a QBA equal to $C$. This is also the same QBA as in the bad-cavity system, \eref{eq:qbanoBAE}. We make the worst-case comparison, and benchmark against all the discussed models. As seen in the figure, the $X$ fluctuations remain well below the expected QBA contributions from all the models. It is hence justified to claim that our measurement truly evades the quantum backaction.


\section{Conclusions}
In this work, we extend the scope of backaction evading (BAE) measurement techniques to a new class of cavity optomechanical systems with large silicon nitride membrane oscillator embedded in a microwave cavity. Our high-conductivity Cu cavity provides an excellent thermal conductivity and allows (i) the mechanical mode to be thermalized near the base temperature of a dilution refrigerator; (ii) nearly eliminate a technical heating of the cavity mode due to the microwave pumping irradiation. In contrast, when the samples are embedded in superconducting Al cavities, the mechanical mode does not thermalize well and the cavity mode heats up significantly. We approach the good-cavity situation and show sideband cooling of a 0.5 mm diameter oscillator close to the ground-state (reaching the mechanical occupation number $n_{m}=0.3$). In this regime, the quantum backaction due to the measurement becomes the main disturbance in the position measurement of a mechanical oscillator.

We then demonstrate that the mechanical quadrature amplitudes can be measured in a continuous and non-demolition manner. We perform BAE measurements on the SiN membrane oscillator by precisely applying two equal-power coherent tones at the mechanical sidebands. We first show a benchmark of the BAE, namely that the spectral linewidth of the measured mechanical $X$ quadrature does not change under different cooperativities. The output photon flux is then shown to grow linearly with increasing the cooperativity in agreement with expectations. The amount of fluctuations, which do not depend on the measurement strength, in the measured quadrature further confirms that our measurement truly evades the quantum backaction, which becomes entirely projected onto the $P$ quadrature. The successful BEA demonstration provides further evidence that large SiN-membrane oscillators coupled to microwave cavities are a viable tool to study mechanical oscillators in the quantum regime.

Looking forward, such large-scale membranes can accommodate milligram-weight objects \cite{Gravity2021}, and the scheme of multiple driving tones can be further extended to prepare these macroscopic objects into the quantum regime using dissipative state preparation. It becomes then possible to measure the gravitational force between non-classical massive oscillators and explore the interface between quantum mechanics and gravity.


\begin{acknowledgments} We acknowledge the facilities and technical support of Otaniemi research infrastructure for Micro and Nanotechnologies (OtaNano) that is part of the European Microkelvin Platform. This work was supported by the Academy of Finland (contracts 307757, 312057), by the European Research Council (101019712), and by the Aalto Centre for Quantum Engineering. The work was performed as part of the Academy of Finland Centre of Excellence program (project 336810). We acknowledge funding from the European Union's Horizon 2020 research and innovation program under grant agreement 824109, the European Microkelvin Platform (EMP). Y.~Liu acknowledges the funding from the National Natural Science Foundation of China (Grant No. 12004044). L. Mercier de Lépinay acknowledges funding from the Strategic Research Council at the Academy of Finland (Grant No. 338565).
\end{acknowledgments}

\appendix
\label{appendix}
\section{Modeling of the system}

This section is intended to present the derivation of the backaction under the different conditions discussed in the main text. In many cases, the derivations are quite straightforward but lengthy, and we thus provide an overall outline.

\subsection{Single-tone driving}

We start from the equations of motion, \eref{eq:eqmotXP}, derived from the Hamiltonian in \eref{eq:Hrotcav}. As a reminder, the cavity mode is transferred to the frame rotating with the pump tone. 

\subsubsection{Bad-cavity, $\omega_m \ll \kappa$}

In the equations of motion, \eref{eq:eqmotXP}, we can approximate $\chi_c = 2/\kappa$. The solution for the mechanics can be written in the form
\begin{subequations}
\label{eq:BadCavCoeff1}
\begin{alignat}{2}
%
%
x & = \tilde{X}_{x} x_{in} + \tilde{X}_{p} p_{in} + \tilde{X}_{ba} X_{c,in}+ \tilde{X}_{baI} X_{c,inI} \,, \\
%
p & = \tilde{P}_{p} p_{in} + \tilde{P}_{x} x_{in} + \tilde{P}_{ba} X_{in}+ \tilde{P}_{baI} X_{c,inI} \,.
\end{alignat}
\end{subequations}
The transduction coefficients that are not explicitly written down are equal to zero. 
%
%
%
The result is
\begin{subequations}
\begin{alignat}{2}
\tilde{X}_{x} & = \tilde{P}_{p} = i \sqrt{\gamma} \omega  \chi_m' \,, \\
\tilde{X}_{p} & = -\tilde{P}_{x}= \sqrt{\gamma} \omega_m \chi_m' \,,  \\
\tilde{X}_{ba} &  = - \sqrt{4 C \gamma} \beta \omega_m \chi_m' \,, \\
\tilde{X}_{baI} &  = - \sqrt{4 C \gamma} \alpha \omega_m \chi_m' \,, \\
\tilde{P}_{ba} & = i \sqrt{4 C \gamma} \beta \omega  \chi_m' \,, \\
\tilde{P}_{baI} & = i \sqrt{4 C \gamma} \alpha \omega  \chi_m'\,.
\end{alignat}
\end{subequations}
Out of these, all other than $X_{ba(I)}, P_{ba(I)}$ are those of non-interacting oscillators. Here, we defined
\begin{equation}
\begin{split}
\label{eq:chidotdef}
\chi_m' & = \frac{1}{(\frac{\gamma}{2} - i \omega )^2 + \omega_m^2} \,,
\end{split}
\end{equation}
and
\begin{equation}
\begin{split}
\alpha & =  \sqrt{\frac{\kappa_i}{\kappa}}  \,,\\
\beta & = \sqrt{\frac{\kappa_e}{\kappa}} \,.
\end{split}
\end{equation}
For evaluating the spectra, we define the following quantities:
\begin{equation}
\label{eq:S0badcav}
\begin{split}
& \mathcal{L}_{\pm}= \frac{1}{(\omega \mp \omega_m)^2 + (\frac{\gamma}{2})^2 } \,,\\
& S_0= \frac{\gamma}{2} \left( \mathcal{L}_+ + \mathcal{L}_- \right) \,. \\
\end{split}
\end{equation}
$S_0$ is a double Lorentzian, peaking at the positive and negative mechanical frequencies.

The following holds for the absolute value of the quantity in \eref{eq:chidotdef}:
\begin{equation}
\begin{split}
& |\chi_m'|^2 = \frac{\mathcal{L}_+ + \mathcal{L}_-}{4 \omega_m^2} = \frac{1}{2 \omega_m^2 \gamma} S_0 \,.\\
\end{split}
\end{equation}

The spectral density of $x$ can be written as
\begin{equation}
\begin{split}
S_x = \langle x(\omega) x(-\omega) \rangle =  S_x^0 +  S_x^T+ S_x^{ba} \,.
\end{split}
\end{equation}
Here, the terms of the rhs are the spectra divided into contributions by the vacuum noise, oscillator thermal occupation, cavity external vacuum noise, and oscillator internal thermal occupation. These are explicitly given by
\begin{subequations}
\begin{alignat}{2}
S_x^0 & =\frac{\gamma}{2} \mathcal{L}_+  \,, \\
S_x^T & = S_0 n_m^T \,, \\
S_x^{ba} & = C \left(1 + 2 n_c^T \right) S_0\,.
\end{alignat}
\end{subequations}
The vacuum noise spectrum exhibits a peak only at positive frequencies.

The fluctuation energy in $x$ becomes
\begin{equation}
\begin{split}
& \langle x^2 \rangle = \frac{1}{2\pi}\int d\omega S_x = \puoli + n_m^T + C(1+2n_c^T) \,,
\end{split}
\end{equation}
which is the result stated in \eref{eq:xpnoBAE}. The spectral density and energy in $p$ are the same as those of $x$.

\subsubsection{Good-cavity, $\omega_m \gg \kappa$}

We assume the sideband cooling conditions: $\Delta = -\omega_m$. Consequently, in contrast to \eref{eq:BadCavCoeff1}, the dynamical quantities depend on the inputs to all others.

For the mechanics, we write
\begin{equation}
\begin{split}
x & = \tilde{X}_{x} x_{in} + \tilde{X}_{p} p_{in} +\tilde{X}_{ba} X_{c,in} + \\
& \tilde{X}_{baI} X_{c,inI} + \tilde{X}_{baP} P_{c,in}+ \tilde{X}_{baPI} P_{c,inI} \,,\\
p & = \tilde{P}_{x} x_{in} + \tilde{P}_{p} p_{in} +\tilde{P}_{ba} X_{c,in} + \\
& \tilde{P}_{baI} X_{c,inI} + \tilde{P}_{baP} P_{c,in}+ \tilde{P}_{baPI} P_{c,inI}\,.
\end{split}
\end{equation}
The backaction terms to the position, as an example, become
\begin{equation}
\begin{split}
\tilde{X}_{ba} & = -\frac{2iG \beta \omega_m}{\sqrt{\kappa}} \chi_e' \,, \\ 
\tilde{X}_{baI} & = -\frac{2iG \alpha \omega_m}{\sqrt{\kappa}} \chi_e' \,,\\ 
\tilde{X}_{baP} & = \frac{2G \beta \omega}{\sqrt{\kappa}} \chi_e' \,,\\ 
\tilde{X}_{baPI} & =  \frac{2G \alpha \omega}{\sqrt{\kappa}} \chi_e' \,,
 \end{split}
\end{equation}
Here, in contrast to \eref{eq:chidotdef}, the susceptibilities exhibit the effective mechanical linewidth, \eref{eq:gammaeff}:
\begin{equation}
\begin{split}
\label{eq:chieffdef}
\chi_e' & = \frac{1}{(\frac{\gamma_{\m{eff}}}{2} - i \omega )^2 + \omega_m^2} \,.
\end{split}
\end{equation}
The energy in the position and momentum is evaluated in the same manner as in the previous section, obtaining \eref{eq:xvarGoodCav}.

\subsection{Bichromatic BAE driving, good-cavity $\omega_m \gg \kappa$}

The equations of motion for the quadratures, based on \eref{eq:hBAE}, are written as
\begin{equation}
\begin{split}
&\dot{X_c} =  -\frac{\kappa}{2} X + \sqrt{\kappa_e} X_{c,in} + \sqrt{\kappa_I} X_{c,inI} \,, \\
&\dot{P_c}  =  2 G X - \frac{\kappa}{2} P_c + \sqrt{\kappa_e} P_{c,in} + \sqrt{\kappa_i} P_{c,inI} \,, \\
&\dot{X} =   - \frac{\gamma}{2} X + \sqrt{\gamma} X_{in}\,, \\  
&\dot{P} = -2 G X_c - \frac{\gamma}{2} P  + \sqrt{\gamma} P_{in}\,. \\  
\end{split}
\end{equation}

Since the system is defined in the frame rotating with the mechanics, the mechanical spectrum now consists of a single Lorentzian at zero frequency. Physically, the two pumps drop sidebands towards the cavity center frequency, where the two sidebands overlap and interfere.

The system dynamics is written as a function of the non-zero transduction coefficients from the inputs, or between the dynamical variables (the term containing $\tilde{P}_{c,x}$):
\begin{equation}
\begin{split}
     & X_c = \tilde{X}_{c} X_{c,in}  + \tilde{X}_{I} X_{c,inI} \,, \\
     & P_c = \tilde{P}_{c,x} X  + \tilde{P}_{c} P_{c,in}+ \tilde{P}_{cI} P_{c,inI} \,, \\
     & X = \tilde{X}_{x} X_{in}  \\
     & P = \tilde{P}_{p} P_{in}  + \tilde{P}_{ba} X_{c,in} + \tilde{P}_{baI} X_{c,inI}\,,
\end{split}
\end{equation}
where
\begin{equation}
\begin{split}
     & \tilde{X}_{x} = \tilde{P}_{p} = \sqrt{\gamma} \chi_m  \,, \\
     & \tilde{P}_{ba} =   -2G  \sqrt{\kappa_e} \chi_m \chi_c \,, \\
     & \tilde{P}_{baI} =   -2G  \sqrt{\kappa_i} \chi_m \chi_c \,, \\
     & \tilde{P}_{c,x} = 2 G \chi_c \,.
\end{split}
\end{equation}
As in the other cases, we suppose $\chi_c = 2/kappa$. One now easily arrives at the mechanical spectra and energy stated in Eqs.~(\ref{eq:BAESQSP},\ref{eq:XPvarBAE}).

We now proceed with the output field, with only the $P_c$ quadrature containing any information on the mechanics:
\begin{equation}
\begin{split}
& P_{c,out} =  \sqrt{\kappa_e} P_c -  P_{c,in} = \\
& \sqrt{\kappa_e} \chi_c [2 G X + \sqrt{\kappa_e}P_{c,in} + \sqrt{\kappa_i} P_{c,inI}] - P_{c,in}
\end{split}
\end{equation}

The output spectrum right after the device becomes
\begin{equation}
\begin{split}
& \langle P_{c,out}  [\omega] P_{c,out} [-\omega] \rangle  \equiv S_{\m{out}}[\omega] = \\ & 16 G^2 \frac{\kappa_e}{\kappa^2} S_{X}[\omega] 
+ \frac{4 \kappa_e}{\kappa} n_c^T +  S_{in}[\omega] \,,
\end{split}
\end{equation}
which repeats \eref{eq:SoutBAE}.

\section{Experimental calibrations}

\subsection{Single tone}

Following the temperature sweep, we calibrate the effective interaction strength $G$ as a function of the generator power setting $P$. The cavity pump photon number $n_c$ is proportional to the power, which relates to $G$ according to \eref{eq:G}. However, a large fraction of the microwave power is lost on its way to the sample. The power loss is
described by the calibration constant $\mathcal{J}$, such that the gain defined in \eref{eq:G} is written as
\begin{equation}
G^2 = \mathcal{J} P\,.
\label{eq:Gmcalpump}
\end{equation}
The quantity that is the most straightforward to measure is the optical damping rate, \eref{eq:gammaopt}, which together with \eref{eq:Gmcalpump} becomes
\begin{equation}
\gamma_{\m{opt}} = \frac{4 \mathcal{J} P}{ \kappa}\,. 
\label{eq:Goptcal}
\end{equation}
We perform a linear fit to the experimentally determined $\gamma_{\m{opt}}$ as a function of the generator power, obtaining 
\begin{equation}
\gamma_{\m{opt}} = \mathcal{L} P\,,
\label{eq:Lcal}
\end{equation}
and comparison with Eq.~(\ref{eq:Goptcal}) yields the calibration coefficient:
\begin{equation}
\begin{split}
&\mathcal{J} = \frac{\mathcal{L}  \kappa}{4 } \,.
\end{split}
\end{equation}
The cooperativity is obtained similarly:
\begin{equation}
\begin{split}
C = & \frac{\gamma_{\m{opt}}}{\gamma}=
\frac{\mathcal{L}}{ \gamma} P \,.
\label{eq:cooper}
\end{split}
\end{equation}

\subsection{Two-tone BAE}

The output photon flux $n_\m{out}$ in \eref{eq:noutBAE} is that right at the cavity output port. The flux $n^d_\m{out}$ finally measured at room temperature is a result of amplification, and some unknown attenuation within the cabling. These are lumped into a power gain $A \gg 1$. The measured flux is therefore
\begin{equation}
\label{eq:ndoutBAE}
\begin{split}
n^d_\m{out}= A \frac{4C  \gamma \kappa_e}{\kappa}  \langle X^2 \rangle
 = A \frac{4C  \gamma \kappa_e}{\kappa} \frac{\gamma k_B }{\gamma_\m{eff} \hbar \omega_0} T \,. 
\end{split}
\end{equation}
The flux is fitted linearly to the high-temperature data points where a good thermalization is obtained:
\begin{equation}
\begin{split}
n^d_\m{out} & = \mathcal{H}  T = \mathcal{H}  \frac{n_m^T \hbar \omega_m}{k_B} \,,
\end{split}
\end{equation}
where $\mathcal{H}$ is the calibration coefficient to be obtained from the fit. This is inverted to yield the bath temperature:
\begin{equation}
\label{eq:nmTBAEcal}
\begin{split}
 n_m^T & =\frac{ n^d_\m{out}}{\mathcal{H}} \frac{k_B}{\hbar \omega_m} \,.
\end{split}
\end{equation}

To infer the mode temperature during the BAE power sweep, we proceed as follows. 
We use the first form of \eref{eq:ndoutBAE}, and fit the measured flux linearly with the cooperativity:
\begin{equation}
\label{eq:ndoutNcal}
\begin{split}
 n^d_\m{out} = \mathcal{N} C \,,
\end{split}
\end{equation}
where $\mathcal{N}$ is the calibration coefficient. In the ideal BAE case, a linear relation holds between the flux and cooperativity because the measured quadrature is not affected. In the practical case, the linear dependence may well be violated because of technical heating or other issues. If, and when, we find a linear regime at least at the lowest powers where the signal is observable, we know that the mode temperature stays constant in that power regime. The energy of the measured quadrature in this regime is denoted by $\langle X^2 \rangle_0$, and this value is obtained from the sideband cooling [\fref{fig:Tsweep} (b)] at a given cooling cooperativity. Combining Eqs.~(\ref{eq:ndoutBAE},\ref{eq:ndoutNcal}) gives
\begin{equation}
\label{eq:BAEcalResults}
\begin{split}
\mathcal{N} & = A \frac{4 \gamma \kappa_e}{\kappa}  \langle X^2 \rangle_0 \,,\\
n^d_\m{out} & =  \mathcal{N} \frac{\langle X^2 \rangle}{\langle X^2 \rangle_0} C \,,\\
\langle X^2 \rangle & = n^d_\m{out} \langle X^2 \rangle_0 \frac{1}{\mathcal{N} C}  \,.
\end{split}
\end{equation}
%



%

\end{document}